# Scalable wavelength-multiplexing photonic reservoir computing

Rui-Qian Li,[1] Yi-Wei Shen,[1] Bao-De Lin,[1] Jingyi Yu,[1] Xuming He,[1] and Cheng Wang[1,2,a]

**AFFILIATIONS**

[1]School of Information Science and Technology, ShanghaiTech University, Shanghai 201210, China
[3]Shanghai Engineering Research Center of Energy Efficient and Custom AI IC, ShanghaiTech University, Shanghai 201210, China

[a]**Authors to whom correspondence should be addressed:** wangcheng1@shanghaitech.edu.cn

**ABSTRACT**

Photonic reservoir computing (PRC) is a special hardware recurrent neural network, which is featured with fast training speed and low training cost. This work shows a wavelength-multiplexing PRC architecture, taking advantage of the numerous longitudinal modes in a Fabry-Perot semiconductor laser. These modes construct connected physical neurons in parallel, while an optical feedback loop provides interactive virtual neurons in series. We experimentally demonstrate a four-channel wavelength-multiplexing PRC, which runs four times faster than the single-channel case. It is proved that the multiplexing PRC exhibits superior performance on the task of signal equalization in an optical fiber communication link. Particularly, this scheme is highly scalable owing to the rich mode resources in Fabry-Perot lasers.

## I. INTRODUCTION

The rapid development of artificial intelligence requires an enormous amount of computational power, which is very challenging for traditional computers based on the von Neumann architecture. Photonic computing is a promising approach to significantly raise the computational power, owing to the fast speed, low latency, and high energy efficiency of light.[1-3] Photonic reservoir computing (PRC) is a special recurrent neural network, where the neurons are connected with multiple feedback loops.[4,5] In contrast to common recurrent neural networks, weights in the input layer and in the hidden reservoir layers of PRCs are fixed, while only weights in the readout layer require training. Therefore, the training speed of PRCs is fast and the training cost is low. One implementation approach of PRCs is connecting the physical neurons with optical waveguides on a single chip.[6] The operation speed of this kind PRC is fast, and the clock rate reaches more than 10 GHz.

However, the integration of nonlinear neurons is technically challenging[7] and the scale is limited due to the transmission loss of light in optical waveguides.[8] Another approach is employing a time-delay loop together with one physical neuron to produce a large number of virtual neurons.[9] The time-delay PRC architecture is usually implemented by using a semiconductor laser with an optical feedback loop,[10] or by using an optical modulator with an optoelectronic feedback loop.[11,12] This time-delay scheme significantly eases the requirement of massive hardware. Nevertheless, the clock rate of the system is inversely proportional to the number of virtual neurons. Consequently, the clock rate of time-delay PRCs is usually limited to tens of MHz.[13,14] In addition to the above two approaches, there are various other implementation schemes,[15] such as using coupled VCSEL arrays[16] or using a spatial light modulator.[17]

In contrast to electronics, photonics have multiple multiplexing dimensions, including wavelength, polarization, space, orbital angular momentum, etc.[18] In the framework of PRCs, Vatin et al. demonstrated a polarization-multiplexing PRC based on the dual-polarization dynamics of a VCSEL, which could process two tasks in parallel.[19] Sunada and Uchida presented a space-multiplexing PRC based on the complex speckle field in a multimode waveguide.[20] Butschek et al. reported a frequency-multiplexing PRC, where 25 comb lines produced by the phase modulation were used as neurons.[21] In addition, Nguimdo et al. numerically proposed a parallel PRC based on the two directional modes in a ring laser.[22] However, the wavelength division multiplexing (WDM) is the most attractive dimension thanks to the broad optical bandwidth of optoelectronic devices. Indeed, the ITU standard dense WDM grid with 50 GHz spacing includes as many as 80 channels. The WDM dimension has been employed in various photonic computing networks, such as the spiking neural network,[23] the convolutional neural network,[24] and the multilayer perceptron.[25] Surprisingly, the deployment of WDM in PRCs

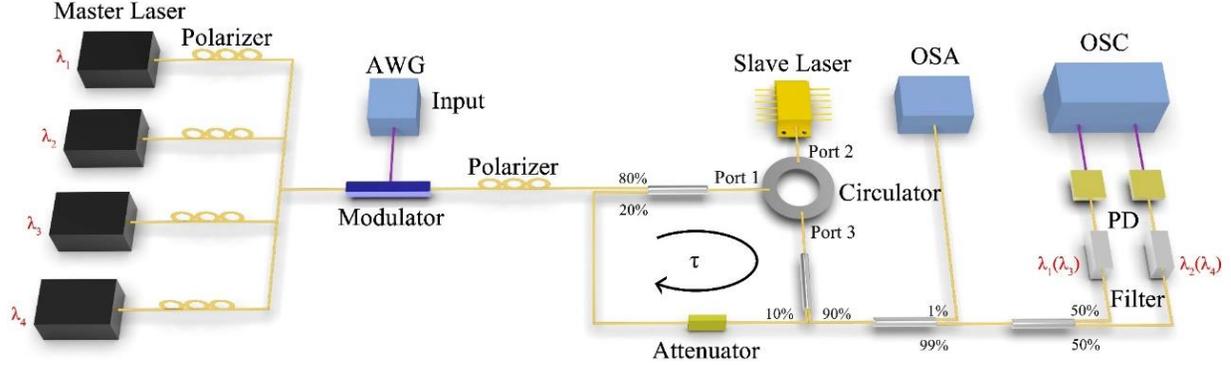

**FIG. 1.** Experimental setup for wavelength-multiplexing PRC. AWG: arbitrary waveform generator; OSA: optical spectrum analyzer; OSC: oscilloscope; PD: photodiode.

has been only discussed in simulations,[26-29] to the best of our knowledge.

This work experimentally presents a wavelength-multiplexing PRC, by using the multiple longitudinal modes in a Fabry-Perot (FP) semiconductor laser. All the modes act as physical neurons, which are connected in parallel through the common gain medium. On the other hand, an optical delay loop is used to produce virtual neurons, which are connected in series. We demonstrate that the four-channel PRC runs four times faster than the single-channel case. In addition, the parallel PRC exhibits better performance on the task of signal equalization in an optical fiber communication link.

## II. WAVELENGTH-MULTIPLEXING SCHEME AND EXPERIMENTAL SETUP

Figure 1 shows the experimental setup for the wavelength-multiplexing PRC architecture. A FP laser with tens of longitudinal modes is used as a slave laser. The modes interact with each other through the common gain medium. All the laser modes are subjected to an optical feedback loop consisting of an optical circulator and two couplers (with ratios of 80:20 and 90:10, respectively). The time-delay loop provides a large number of virtual neurons and constructs the hidden reservoir layer of the PRC.[9,10] The optical feedback strength of the feedback loop is tuned by an optical attenuator. Four single-mode external-cavity lasers are used as the master lasers, and the wavelength ($\lambda_{1-4}$) of each laser is finely tuned to align with one longitudinal mode of the slave laser, respectively. All the master lasers are unidirectionally injected into the slave laser through the optical circulator. The optical injection is operated in the stable locking regime, which is bounded by the Hopf bifurcation and the saddle-node bifurcation.[30,31] After the power amplification with an Erbium-dope fiber amplifier, the polarization of each master laser is aligned with that of the Mach-Zehnder intensity modulator (EOSPACE, 40 GHz bandwidth) by using a polarization controller, respectively. Then, the polarization of the modulated light is re-adjusted to align with that of the slave laser. Every symbol of the input signal under test is firstly multiplied with a mask, which consists of a random binary sequence of {0, 1}.[32] The mask plays a crucial role in maintaining the transient state of the nonlinear laser system, which is the fundamental requirement of time-delay PRCs. In addition, the duration between each bit of the mask determines the temporal interval of virtual neurons. The preprocessed signal is produced from the arbitrary waveform generator (AWG, Keysight, 25 GHz bandwidth). This radio-frequency signal is amplified before driving the intensity modulator. In this way, the signal at the input layer of the PRC is injected into the slave laser at the hidden reservoir layer for nonlinear processing. The optical spectrum of the output signal is measured by an optical spectrum analyzer (OSA, Yokogawa, 0.02 nm resolution bandwidth). At the output layer, the light is split into two branches by using a 50:50 splitter. Each branch analyzes one longitudinal mode through using a bandpass filter (0.95 nm bandwidth), respectively. The two optical signals are detected in parallel by high-speed photodiodes (PD, 25 GHz bandwidth and 50 GHz bandwidth). After power amplification, the temporal waveforms of both channels are recorded on the digital oscilloscope (OSC, Keysight, 59 GHz bandwidth), simultaneously. The two modes with wavelengths of $\lambda_{1,2}$ are recorded firstly, while another two modes with wavelengths of $\lambda_{3,4}$ are tracked in the second-round measurement. It is worthwhile to point out that the four modes can be tracked simultaneously in case a proper wavelength demultiplexer is employed.

In the experiment, the delay time of the optical feedback loop is measured to be about $\tau$=65.3 ns. The time interval of the virtual neurons in the reservoir is $\theta$=0.05 ns, which is governed by the modulation rate of the modulator at 20 Gbps. The number of virtual neurons is set at $N$=80 throughout the experiment. The weights of the output layer in the PRC are trained with the algorithm of ridge regression.[32] The sampling rate of the AWG is set at 60 GSa/s and the rate of the oscilloscope is at 80 GSa/s. For the single channel PRC with only one master laser, the clock cycle of the system is $T_c$=4.0 ns, which is determined by the formula $T_c=\theta \times N$. When the WDM scheme with multiple master lasers is employed, the neuron number of each channel is inversely proportional to the channel number $m$ as $N/m$. Consequently, the clock cycle of the system scales down with the channel number as $T_c=\theta \times N/m$. For the four-channel PRC in Fig. 1, the clock cycle reduces down to $T_c$=1.0 ns, which is four times faster than the one-channel case. It is stressed that the clock cycle $T_c$ of the PRC in Fig. 1 is significantly shorter than the delay time $\tau$, which is different to the common synchronous time-delay PRCs. Our recent work has proved that this asynchronous architecture is beneficial to improve the performance of PRCs,[33] owing to the off-resonance effect.[34]

### III. EXPERIMENTAL RESULTS

In the experiment, the slave FP laser exhibits a lasing threshold of $I_{th}$=8.0 mA at the operation temperature of 20 ℃. The laser is biased at 3.5×$I_{th}$ with an output power of 3.2 mW, unless stated otherwise. The peak of the optical spectrum in Fig. 2(a) locates around 1545 nm, with a free spectral range of about 1.23 nm. When the slave laser is subject to the optical injection from one master laser in Fig. 2(b), only the injected mode keeps lasing, while other longitudinal modes are suppressed due to the gain reduction of the laser medium.[31] When two master lasers respectively lock two modes in Fig. 2(c), only the two injected modes remain lasing. In the same way, Fig. 2(d) shows that four modes keep lasing when four master lasers inject into the slave laser, simultaneously. It is noted that all the modes in the FP laser interact with each other due to the cross-gain coupling effect, rather than emit independently.[29] Therefore, the four modes in Fig. 2(d) act as connected physical neurons in parallel. In addition, every mode subject to the feedback loop in Fig. 1 produces a reservoir of virtual neurons, respectively. As a result, the four modes in the same gain medium generate four connected reservoirs together with a large number of virtual neurons.

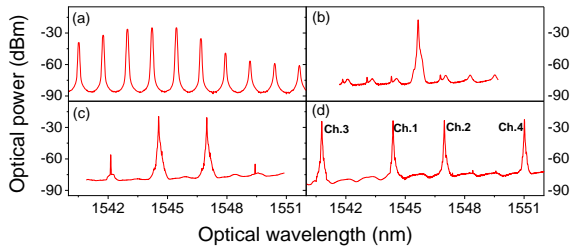

**FIG. 2.** Optical spectra of the slave FP laser under the operation of (a) free running, (b) one-channel injection, (c) two-channel injection, and (d) four-channel injection. The label of the channels is marked in (d).

The performance of the wavelength-multiplexing PRC is tested on the task of signal equalization in an optical fiber communication link.[35] The optical signal at the receiver side is distorted due to the effect of chromatic dispersion and the effect of Kerr nonlinearity in optical fibers. The aim of the task is to recover the original signal at the transmitter based on the distorted one at the receiver. It has been widely shown that various digital artificial neural networks could well compensate the linear dispersion and the nonlinear impairment in both intensity modulation and direct detection (IMDD) communication links and coherent communication links.[36] This work takes into account an IMDD link described by the nonlinear Schrödinger equation,[37]

$$\frac{\partial E}{\partial z} + \frac{\alpha}{2} E + j \frac{\beta_2}{2} \frac{\partial^2 E}{\partial t^2} = j\gamma |E|^2 E, \quad (1)$$

where $E(x,t)$ is the slowly-varying envelop of the electric field in the fiber. The attenuation coefficient of the fiber is α=0.2 dB/km, the chromatic dispersion coefficient is $\beta_2$=-21.4 ps$^2$/km, and the Kerr nonlinearity coefficient is γ=1.2 /(W·km).[38] The signal is modulated by the non-return-to-zero for-mat with random bits of {0, 1}. The modulation rate of the, signal is 25 Gbps, the transmission length is 50 km, and the launch power is 4 mW, unless stated otherwise. 30000 symbols are used to train the PRC, and 15000 symbols are used to test the performance. The performance is quantified by the bit error rate (BER). The measurement is repeated four rounds, and the mean BER and the standard deviation of uncertainty are collected.

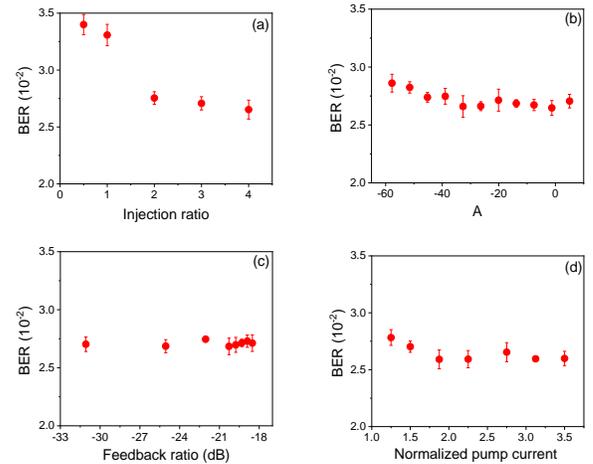

**FIG. 3.** Performance of the single-channel PRC. Effects of (a) the injection ratio $R_{inj}$, (b) the detuning frequency $\Delta f_{inj}$, (c) the feedback ratio $R_{ext}$, and (d) the normalized pump current $I/I_{th}$. The default operation conditions are $R_{inj}$=4.0; $\Delta f_{inj}$: near Hopf bifurcation; $R_{ext}$=-30.3 dB; $I/I_{th}$=3.5.

We first investigate the performance of the single-channel PRC. Figure 3(a) shows that the BER declines nonlinearly with the injection ratio, owing to the increased signal-to-noise ratio of the PRC system. In addition, increasing the detuning frequency in Fig. 3(b) from the side of the saddle-node bifurcation to the side of Hopf bifurcation reduces the BER. That is, the optimal PRC performance is achieved in the vicinity of the Hopf bifurcation. This is because the positive frequency detuning of optical injection reduces the damping factor of the slave laser, leading to richer dynamics of virtual neurons.[39,40] Figure 3(c) shows that the BER of the signal is insensitive to the optical feedback ratio. It is remarked that the PRC is always operated in the stable regime of optical feedback. The upper limit of the stable regime is bounded by the critical feedback level, beyond which the slave laser becomes unstable.[30,41] The critical feedback level of the slave laser without optical injection is measured to be about -19.3 dB. However, our recent work found that the optical injection significantly raised the critical feedback level of the slave laser.[33] Figure 3(d) shows that the BER firstly decreases with increasing pump current, and then saturates around 0.026 when the pump current is larger than $2.0 \times I_{th}$. Interestingly, the impacts of the above four operation parameters on the signal equalization task are similar to those on the prediction task of Santa Fe chaos in.[33]

the signal distortion. The average BER of the one-channel PRC (squares) is 0.028. In comparison, the average BER of both the two-channel PRC (triangles) and the four-channel PRC (dots) is 0.024, which is 14% smaller than the one-channel case. That is, the WDM scheme improves the PRC performance on the signal equalization task. This is because the laser mode interaction provides parallel connections of the virtual neurons, in addition to the series connections arising from the optical feedback loop. Besides, we remind that the clock rate (1.0 GHz) of the four-channel PRC is four times faster than that of the one-channel case (0.25 GHz), taking advantage of the WDM architecture. According to Fig. 3(a), the performance of the four-channel PRC can be further improved by raising the injection ratio to 4.0 (like one- and two-channel cases) instead of only 1.0.

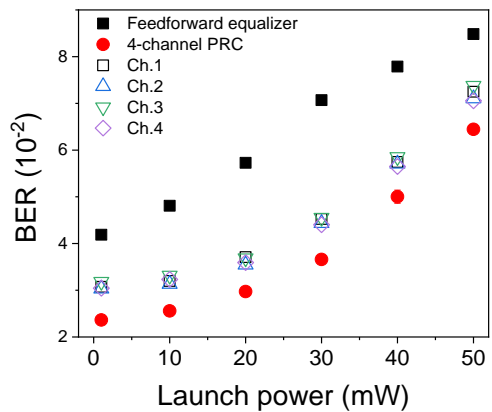

FIG. 5. Performance comparison between the four-channel PRC (dots) and the feedforward equalizer (squares) for a broad range of launch power. The open symbols represent the BERs of each channel, respectively.

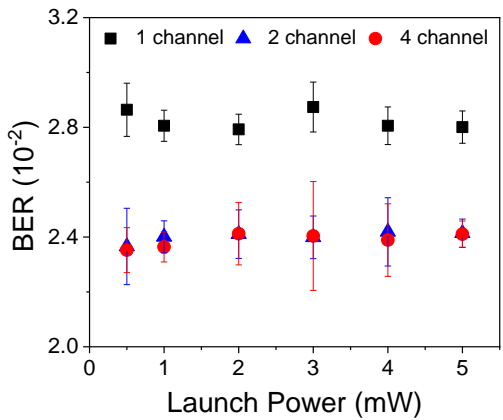

FIG. 4. Performances of the wavelength-multiplexing PRCs versus the launch power of the transmitted signal. The feedback ratio is $R_{ext}$=-30.3 dB and the pump current is $I/I_{th}$=3.5. The optical injection conditions are: $R_{inj}$=4.0, $\Delta f_{inj}$= -20.1 GHz for the one-channel case (squares); $R_{inj}^{1,2}$=4.0, $\Delta f_{inj}^1$ = -34.8 GHz, $\Delta f_{inj}^2$ = -32.3 GHz for the two-channel case (triangles); $R_{inj}^{1-4}$ =1.0, $\Delta f_{inj}^1$ = -31.1 GHz, $\Delta f_{inj}^2$ = -23.6 GHz, $\Delta f_{inj}^3$ = -36.0 GHz, $\Delta f_{inj}^4$ = -77.0 GHz for the four-channel case (dots).

Figure 4 compares the performances between PRCs with different number of channels. It is shown that all the three PRCs are insensitive to the launch power of the transmitted signal as long as the power is less than 5.0 mW. This is because the small launch power does not stimulate strong Kerr nonlinearity, and hence the chromatic dispersion dominates

In order to investigate the ability of the PRC for the compensation of fiber nonlinearity, we artificially raise the launch power of the transmitted signal up to 50 mW in Fig. 5. It is shown that the BER of the four-channel PRC increases nonlinearly with the launch power from 0.024 at 1.0 mW to 0.064 at 50 mW. In addition, the BER of every channel (open symbols) with a neuron number of 20 rises nonlinearly as well. Obviously, the four-channel PRC with a total neuron number of 80 performs better than each channel, because more neuron dynamics are involved. The performance of the PRC is compared with the feedforward equalizer, which is a transversal filter that linearly combines the received symbol and its neighbors.[35] That is, the feedforward equalizer only compensates the chromatic dispersion effect. Figure 5 shows that the BER of the feedforward equalizer (tap number is 5) increases almost linearly with the launch power from 0.042 at 1.0 mW to 0.085 at 50 mW. In comparison, the PRC exhibits better performance at both low and high launch powers. On one hand, this is because the PRC has fading memory effect owing to the nature of recurrent neural networks. Therefore, the PRC can better compensate the distortion of chromatic dispersion. On the other hand, the PRC is a typical nonlinear system and thereby can compensate the distortion of Kerr nonlinearity as well. This comparison result is in agreement with those observed in literatures.[42-44]

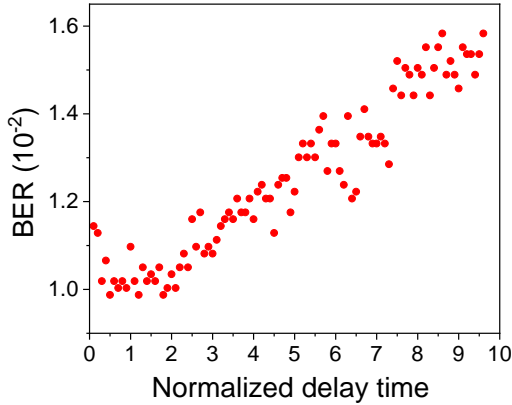

**FIG. 6.** Simulated effect of the normalized delay time $\tau/T_c$ on the PRC performance.

### IV. DISCUSSION

In the above experiment, the delay time of the feedback loop is fixed at $\tau$=65.3 ns without any optimization. Although optimization of the PRC performance is beyond the scope of this work, this section discusses the effect of delay time in simulation so as to provide some insight on future experiment design. We simulate the PRC using the model described in.[29] The slave laser is described by the rate equation approach, which takes into account the dynamics of the carriers, the photon, and the phase of the electric field. The optical feedback effect and the optical injection effect are described by the classical Lang-Kobayashi model.[45,46] In the simulation, the neuron number of the single channel PRC is set at 80. The neuron interval is set at 0.01 ns, and hence the clock cycle is $T_c$=0.8 ns. The simulated BER in Fig. 6 firstly goes down with the normalized delay time starting from $\tau/T_c$=0.1. The optimal PRC performance is achieved within the normalized time range of 0.5 to 2.0, and the best BER is around 0.010. In comparison, the optimal time range for the prediction task of Santa Fe chaos is from 2.0 to 4.0.[33] Interestingly, the BER jumps up to 0.011 at $\tau/T_c$=1.0, where the delay time is synchronous with the clock cycle. The performance degradation is attributed to the detrimental resonance effect.[33,34] For $\tau/T_c$>2.0, the BER almost rises linearly with increasing delay time, and its value reaches 0.016 at $\tau/T_c$=9.6. In the experimental setup of Fig. 1, nevertheless, the feedback delay time is more than an order of magnitude than the clock cycle, which is far away from the optimal value. Consequently, future experiment requires an optimization of the delay time to achieve the best signal equalization performance.

### V. CONCLUSION

In summary, we have experimentally demonstrated a wavelength-multiplexing PRC based on the numerous longitudinal modes in a FP laser. The modes play the role of connected physical neurons in parallel. Meanwhile, an optical feedback loop produces virtual neurons, which are connected in series through the time-multiplexing effect. It is shown that the four-channel PRC runs four times faster than the single-channel case, and the clock rate reaches up to 1.0 GHz. It is found that the four-channel PRC exhibits superior performance on the signal equalization task, owing to the interaction of neurons both in parallel and in series. The proposed WDM scheme is highly scalable owing to the rich mode resources in FP lasers. Future work will scale up the number of WDM channels and further raise the clock rate of the PRC.

### ACKNOWLEDGMENTS

This work was funded by the Shanghai Natural Science Foundation (No. 20ZR1436500).

### AUTHOR DECLARATIONS

**Conflict of Interest**

The authors have no conflicts to disclose.

**Author contributions**

R.Q.L. and Y.W.S. contributed equally to this work.

**Rui-Qian Li**: Data curation (equal); Investigation (equal); Validation (equal); Visualization (equal); Writing–original draft (equal). **Yi-Wei Shen**: Data curation (equal); Investigation (equal); Validation (equal); Visualization (equal); Writing–original draft (equal). **Bao-De Lin**: Data curation (equal); Investigation (equal); Validation (equal); Visualization (equal); Writing–original draft (equal). **Jingyi Yu**: Methodology (equal); Project administration (equal); Supervision (equal); Writing – review & editing (equal). **Xuming He**: Methodology (equal); Project administration (equal); Supervision (equal); Writing – review & editing (equal). **Cheng Wang**: Methodology (lead); Project administration (lead); Supervision (equal); Funding acquisition (lead); Writing – review & editing (lead).

### DATA AVAILABILITY

The data that support the findings of this study are openly available in
https://zenodo.org/record/7961785#.ZGx_FXZByHu